# Emergency Resource Layout with Multiple Objectives under Complex Disaster Scenarios

Changwei Yuan[1] and Shengxuan Ding[2 *] and Yin Wang[3] and Jian Feng[4] and Ningyuan Ma[5]

[1] School of transportation engineering, Chang'an University, Xi'an, 710061; PH 139-9280-2916; E-Mail: changwei@chd.edu.cn

[2] (Corresponding Author) School of transportation engineering, Chang'an University, Xi'an,710061; PH 184-0519-6669; E-Mail: 2020234075@chd.edu.cn

[3] School of transportation engineering, Chang'an University, Xi'an, 710061; PH 156-9182-7809; E-Mail: wyin@chd.edu.cn

[4] School of transportation engineering, Chang'an University, Xi'an, 710061; PH 136-3021-6091; E-Mail: 2021234097@chd.edu.cn

[5] School of transportation engineering, Chang'an University, Xi'an, 710061; PH 187-0297-1922; E-Mail: 2020034025@chd.edu.cn

# Abstract

Effective placement of emergency rescue resources, particularly with joint suppliers in complex disaster scenarios, is crucial for ensuring the reliability, efficiency, and quality of emergency rescue activities. However, limited research has considered the interaction between different disasters and material classification, which are highly vital to the emergency rescue. This study provides a novel and practical framework for reliable strategies of emergency rescue under complex disaster scenarios. The study employs a scenario-based approach to represent complex disasters, such as earthquakes, mudslides, floods, and their interactions. In optimizing the placement of emergency resources, the study considers government-owned suppliers, framework agreement suppliers, and existing suppliers collectively supporting emergency rescue materials. To determine the selection of joint suppliers and their corresponding optimal material quantities under complex disaster scenarios, the research proposes a multi-objective model that integrates cost, fairness, emergency efficiency, and uncertainty into a facility location problem. Finally, the study develops an NSGA-II-XGB algorithm to solve a disaster-prone province example and verify the feasibility and effectiveness of the proposed multi-objective model and solution methods. The results show that the methodology proposed in this paper can greatly reduce emergency costs, rescue time, and the difference between demand and suppliers while maximizing the coverage of rescue resources. More importantly, it can optimize the scale of resources by determining the location and number of materials provided by joint suppliers for various kinds of disasters simultaneously. This research represents a promising step towards making informed configuration decisions in emergency rescue work.

**Key words:** Emergency resource, Multiple types of materials, Complex disaster scenarios, Multiple rescue subjects, Multi-objective optimization, NSGA-II-XGB algorithm

# 1.Introduction

The occurrence of major disasters such as extreme weather, floods, and geological events have resulted in significant loss of life and property [1]. Due to the involvement of multiple rescue departments, the unpredictable nature of disasters, and the uneven distribution of emergency supplies, there is a critical need for efficient storage and location selection of emergency materials[2]. Current research is focused on developing multi-agent trust systems[3],information sharing and coordination[4], and linked storage of materials and equipment[5]. However, the current classification standards for emergency supplies are limited[6] and coordinated rescue efforts for multi-disaster scenarios are weak[7]. While there has been research on allocating emergency resources during the response stage[8-10], insufficient attention has been given to resource location planning during the preparation stage[11]. This is crucial for ensuring reliable emergency responses. There are several challenges that need to be addressed in the process of selecting emergency rescue sites.

Firstly, in uncertain scenarios with complex data, the integration of information is often delayed[12], and the evolving nature of disasters poses difficulties for storing and allocating emergency supplies[13].Therefore, this research focuses on integrating multidimensional disaster information and considering how the evolution of disasters impacts emergency material reserve strategies.

Secondly, the distribution of equipment is uneven, with limited types and insufficient data sharing for supporting reserves[14]. This often leads to mismatches between supply and demand, and delays in transportation when allocating emergency materials across regions, resulting in failure to deliver emergency materials to the disaster area promptly, and delaying the best time for rescue[15]. The existing classification system for emergency supplies, based on their uses, is inadequate for cross-regional allocation. Therefore, it is urgent to solve the problem of reserve network layout and reserve setting and find a more detailed multi-material collaborative reserve strategy.

Thirdly, multi-link coordination for disaster rescue is weak, with difficulties in sharing emergency resources and low utilization rates[16]. An emergency linkage mechanism needs to be implemented to store and allocate emergency resources in a unified manner, and enable collaborative work among emergency departments.

To address these issues, this research aims to integrate multi-dimensional disaster information and consider the impact of disaster evolution on emergency material reserve strategies. The study will also focus on the distribution of emergency equipment and develop a more detailed multi-material collaborative reserve strategy to enable cross-regional allocation of emergency supplies. Moreover, the study will investigate an emergency linkage mechanism to promote collaboration and coordination among different emergency departments. The proposed approach combines multi-subject suppliers to make unified layout decisions on emergency reserves from a global perspective.

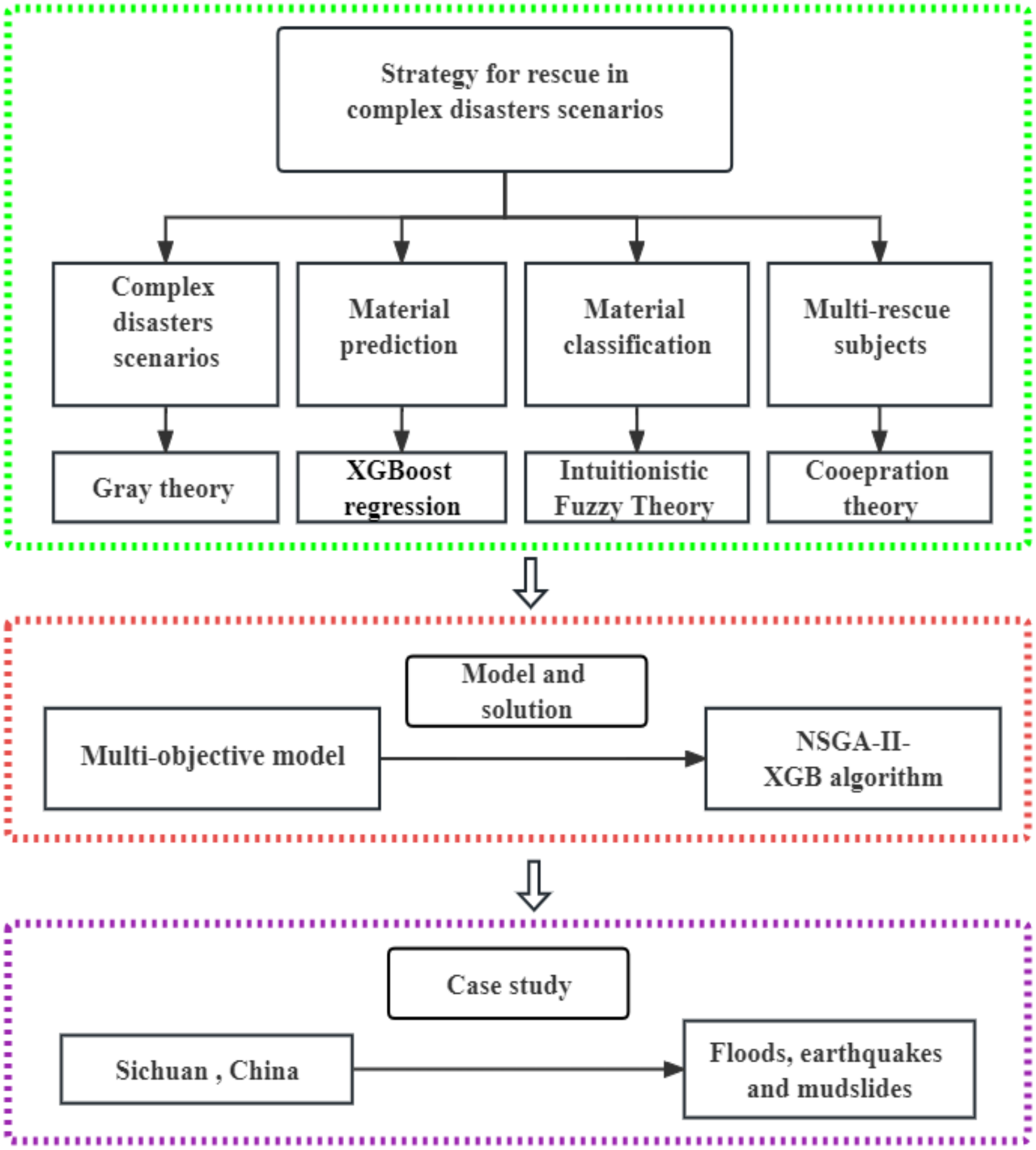

**Figure.1 Framework of this paper**

In the next section, we motivate our approach by reviewing prior related research. Thereafter, we present the structure of our methods including material prediction and classification, multi-objective model and solution approach. In Section 4, we describe the dataset, and subsequently present the result of methodology to verify the proposed model. We then continue provide discussion and concluding remarks in Section 5. The framework of this paper is shown in Figure1.

# 2. Literature review

## 2.1 Multiple rescue models

The fundamental framework for selecting an emergency rescue site comprises the P-median model, P-center model, and set coverage model. Scholars have further developed the basic model by introducing deterministic models and other variations.

In the field of earthquake emergency relief planning, various models have been proposed to determine the optimal location of relief facilities, pre-storage and response phases, and rescue dispatch strategies while considering uncertainties such as demand, supply, cost, facility failure, traffic congestion, and queuing delays. XG A[17]developed a multi-objective location model for earthquake emergency relief materials, which divided the coverage threshold based on time series characteristics and adopted the target deviation rate minimization model and NSGA-II Algorithm to solve the problem. The feasibility of the model was verified through a case study in Aba Prefecture, Sichuan. M Haghi[18]proposed a multi-objective programming model to determine the location of the rescue cargo distribution center, considering pre-disaster and post-disaster uncertainties and aiming for fair distribution of relief supplies while minimizing the total cost. Xie [19]proposed a scenario-based stochastic mixed integer nonlinear model that integrates facility disruption risk, traffic congestion, and queuing delays into one model, deriving the expected total cost of the system in the fuzzy facility outage scenario and deducing the lower and upper bounds of this problem. Sensitivity analysis highlighted the importance of considering queuing in the facility and the quantity of materials in the location selection. A Dyen[20]established a two-stage stochastic mixed integer linear programming model for rescue site selection, aiming to minimize the total cost of facility siting, inventory holding, transportation, and shortages while determining pre-disaster and post-disaster rescue centers, the number of relief materials in the pre-disaster rescue center, and the shortage of relief materials. A Bozorgi-Amiri[21] developed an eclectic planning model that regards demand, supply, and cost as uncertain parameters in a multi-objective model, maximizing the satisfaction of affected areas and minimizing the maximum shortage sum as a non-dominated compromise solution. Ni[22] established a minimum-maximum robust model to optimize the location of facilities in the disaster relief network, the location of emergency stocks, and the rescue dispatch strategy, capturing uncertainty in the boundary parameters of the constraint conditions. The feasibility of the model was verified through a case study in the 2010 Yushu

County earthquake in Qinghai Province. In terms of multi-party collaborative hierarchical location model, S Tofighi[23]solved a two-tier rescue network design problem involving multiple central warehouses (CWs) and local distribution centers (LDCs), and established a scenario based on two stages- Stochastic programming (SBPSP) method, the first stage determines the location and inventory level of relief supplies, taking into account the uncertainty of supply and demand and the accessibility of the road network after the earthquake; the second stage, formulate rescue distribution plans according to various disaster scenarios , which minimizes the time-weighted, total inventory cost, and weighted shortage cost. JH Zhang[24]considered the joint reserve model of the government and the agreement supplier, represented the uncertainty of demand based on the scene, established a nonlinear mathematical model before and after the disaster to analyze the material storage strategy of the government and the agreement supplier, and sensitively related parameters gender analysis. Sheu JB[25] proposed a multi-objective programming model for global rescue network configuration hierarchical network, considering the potential risk cost, based on integer programming and hierarchical clustering analysis methods, determined the location and quantity of service areas and facilities in the GLS network and coverage, minimize network configuration costs, maximize operating profits and customer satisfaction.

Numerous studies have explored emergency rescue site selection models, which include coverage, hierarchical division, and multi-objective optimization. These studies mainly aim to ensure the reliability and minimize the uncertainty of emergency rescue site selection. However, there is a lack of research that integrates multi-targets from various perspectives, specifically in multi-disaster scenarios and multi-type material emergency rescue site selection.

## 2.2 Multiple types of materials and rescue subjects

In their research on emergency material classification, Hill[26] criticized the current standards as insufficient for practical application. They employed the AHP (analytical hierarchy process) to conduct qualitative and quantitative research and aggregated the results into a classification analysis. Mete[27]focused on the storage and distribution of medical emergency supplies in various disaster scenarios. They developed a stochastic programming model for locating and distributing medical emergency supply facilities, using the example of an earthquake in Seattle to test their approach. X Guan[28] proposed a site selection model for earthquake emergency service bases based on a hybrid multi-attribute decision-making method. Their method considered the mixed uncertainty of location decision information and provided a new weighting method and aggregation process without requiring information transformation. Finally, they proved the feasibility and effectiveness of their method using numerical examples.

The focus of research on emergency site selection involving multiple parties is typically on the emergency response stage. For example, [29, 30]highlights the importance of stakeholders participating in emergency response and coordinated rescue, based on power, legitimacy, and urgency. JH Zhang's research (i) considers a joint reserve model involving the government and an agreement supplier and establishes a nonlinear mathematical model to analyze the material reserve strategy before and after a disaster, along with conducting sensitivity analysis on relevant parameters. In the emergency management preparation stage, [31] transformed the site selection parameters into decision variables in a multi-party collaborative emergency response decision-making model and developed an emergency service facility siting model based on multi-party coordination that aims to minimize completion time and service acquisition difficulty cost. [32] takes a global perspective combined with administrative divisions to examine the layout of site selection under emergency linkage, assesses the impact of emergency material classification results on site selection decision-making schemes, and ultimately comes up with a provincial and municipal reserves site selection decision and resource allocation scheme. [33] targets the characteristics of multi-subjectivity, complexity, and uncertainty of emergency decision-making under unconventional emergencies, and constructs an asymmetric association directed collaborative network to select the emergency plan for unconventional emergencies, based on individual and collaborative information between emergency plans. Previous study also provided a guideline for the application of Open AI in medical rescue[34]. The current research underscores the significance of coordinated rescue in emergency response and analyzes multi-agent rescue in site selection pre-storage strategy in terms of parameter determination, collaborative network, rescue subject classification, and joint reserve of the government and agreement suppliers. However, larger-scale and more complex multi-objective optimization scenarios must be addressed when facing uncertain emergency service needs, such as disaster occurrence time, location, consequences, and road network damage.

The current research suggests a material grading system to classify emergency resources based on fuzzy evaluation analysis and analytic hierarchy process. This system provides a scientific and rational approach to resource allocation. However, the existing index system lacks sufficient secondary indicators and empirical analysis, relying on hypothetical parameters. To improve the system, this paper aims to refine secondary indicators and adjust site selection model parameters using actual data as a reference. When dealing with multiple rescue subjects, it is important to consider the construction of a multi-level network and a model that accounts for uncertainty in multiple rescue subjects. However, in cases of larger and more intricate multi-objective optimization scenarios, the current collaborative research approach among multiple parties may no longer be viable.

### 2.3 Complex disaster scenarios

In multi-hazard research, Rawls C G[5]conducted research on selecting locations for emergency service facilities in the context of natural disasters such as hurricanes, using a two-stage Mixed Integer Programming Siting Model. J Shu[35] developed a rescue and supply network design model using a variant of the expander graph to minimize costs of emergency facilities and supply locations for large-scale disasters such as floods and hurricanes. The proposed model was validated using the 2013 northeast flood case study. G Erbeyolu[36] proposed an emergency rescue location model that considers facility location and inventory level, using the benders method to find the optimal solution. Ginger Y[37] examined the impact of system interruption on emergency logistics for dangerous goods, and proposed a time-varying risk assessment method using a two-stage robust optimization approach. A two-level mixed integer programming model was developed, and the model was accurately solved using column and constraint generation algorithms. The model was tested on random instances of various sizes.

Regarding multi-hazard situations, previous studies have presented different categories and intensities of disaster scenarios and have identified the uncertain decision-making information when facing outages. Nonetheless, there has been limited research on how different types of disasters interact and their simultaneous occurrence.

Numerous scholars have extensively researched emergency rescue site selection models and algorithms, multi-party collaborative site selection, and material classification, among other topics, in the literature. Their findings provide the necessary theoretical foundation for selecting emergency material sites during times of crisis. This study builds upon the existing knowledge by focusing on the emergency rescue site selection model in multi-disaster scenarios. It analyzes the systematization and substitutability of disaster emergency supplies' classification and classification and develops an emergency equipment reserve layout model based on multi-rescue subjects. This research broadens and deepens the understanding of emergency relief supplies' site selection and reserve models, offering theoretical support for future site selection and layout research.

# 3. Methods

Disaster relief requires careful consideration of emergency resource allocation to minimize casualties and economic losses. In recent years, inadequate resource allocation, insufficient preparation, and secondary disasters have made rescue operations more difficult. To achieve effective disaster relief, it is crucial to study the optimal scale and location of emergency resources.

## 3.1 Rescue strategy under complex disasters scenarios

### 3.1.1 Coupling degree of complex disasters

This paper employs the disaster risk coupling matrix technique to perform a comprehensive analysis of the consequences of various hazards. Despite the presence of multiple risks, the study focuses on examining the interdependent relationship among earthquakes, typhoons, and floods. The investigation covers diverse aspects such as the nature and development of disasters, the rescue team involved and their necessary reserves, and the utilization of location selection models.

The gray theory is applied in this research to assess the level of interdependence among various disasters[38]. A rating scale ranging from 1 to 9 is employed to construct a decision matrix and evaluate the relative significance of each indicator. To begin with, the weight of each index is computed by multiplying the row index values of the matrix.

$$\mathrm{M}_i = \prod_{\mathrm{i}=1}^{\mathrm{n}} \mathrm{x_{ij}} (\mathrm{i}=1,2,...,\mathrm{n}) \tag{3.1}$$

Where, $\mathrm{x_{ij}}$ is the weight of evaluate index i for disaster j.

Calculate the nth root of M:

$$\overline{\mathrm{W}_i} = \sqrt[\mathrm{n}]{\mathrm{M}_i} \tag{3.2}$$

Determine the weight matrix:

$$\mathrm{Wi} = \overline{\mathrm{W}_i} / \sum_{\mathrm{i}=1}^{\mathrm{n}} \sqrt[\mathrm{n}]{\overline{\mathrm{W}_i}} \tag{3.3}$$

$$\lambda_{max} = \frac{1}{n} \sum_{\mathrm{i}=1}^{\mathrm{n}} \frac{(\mathrm{BW})_i}{\mathrm{W}_i} \tag{3.4}$$

$$\mathrm{BW} = \begin{bmatrix} \mathrm{x_{11}} & \cdots & \mathrm{x_{1n}} \\ \vdots & \ddots & \vdots \\ \mathrm{x_{m1}} & \cdots & \mathrm{x_{mn}} \end{bmatrix} \begin{bmatrix} \mathrm{W_1} \\ \vdots \\ \mathrm{W_n} \end{bmatrix} = ({d_{i1}}^{max}, {d_{i2}}^{max}, \dots, {d_{in}}^{max})^T \tag{3.5}$$

Where, ${d_{i1}}^{max}$ is the optimal vector of the secondary coupling index.

Consistency check:

$$CR=\frac{\lambda_{max}-n}{(n-1)RI} \tag{3.6}$$

Where, CR is the random consistency ratio, RI is the random consistency index.

The correlation calculation formula is as follows:

$$\lambda_{ij}=\frac{\min|d_{i1}-d_{ik}{}^{max}|+\rho\max|d_{ik}-d_{ik}{}^{max}|}{|d_{i1}-d_{ik}{}^{max}|+\rho\max|d_{ik}-d_{ik}{}^{max}|} \tag{3.7}$$

Where, $\rho$ represents the resolution coefficient, usually selects 0.5.

The coupling degree evaluation matrix is as follows:

$$P_i=(\lambda_{i1},\lambda_{i2},\ldots,\lambda_{in})^T \tag{3.8}$$

Calculate the composite index of the coupling index:

$$X_i=W^T P_i \tag{3.9}$$

Where, $X_i$ represents the comprehensive index of the secondary index; $W^T$ is the feature vector.

For a comprehensive multi-hazard coupled system, the efficiency function of its coupling index can be expressed as:

$$\mu_{ij}=\frac{X_{ij}-\beta_{ij}}{\alpha_{ij}-\beta_{ij}}, X_{ij} \text{ is positive} \tag{3.10}$$

$$\mu_{ij}=\frac{\alpha_{ij}-X_{ij}}{\alpha_{ij}-\beta_{ij}}, X_{ij} \text{ is negative} \tag{3.11}$$

The values in this paper are as follows: $\alpha_{ij}$=1, $\beta_{ij}$=0.

The multiple system coupling formula is as follows:

$$C=\sqrt[m]{\frac{U_1U_{2\ldots}U_m}{\prod_{j=1}^{m}\prod_{i=1, i\neq j}^{m}(U_i+U_j)}} \tag{3.12}$$

### 3.1.2 Material prediction

The prediction of demand for emergency rescue materials is a crucial aspect of disaster response, as it enhances the efficiency of emergency relief and ensures the survival of disaster victims. In recent years, the XGboost algorithm has gained popularity as a machine learning algorithm and has proven effective in predicting the demand for emergency relief supplies[39]. This article will discuss the selection criteria for utilizing the XGboost algorithm in forecasting the demand for emergency relief materials.

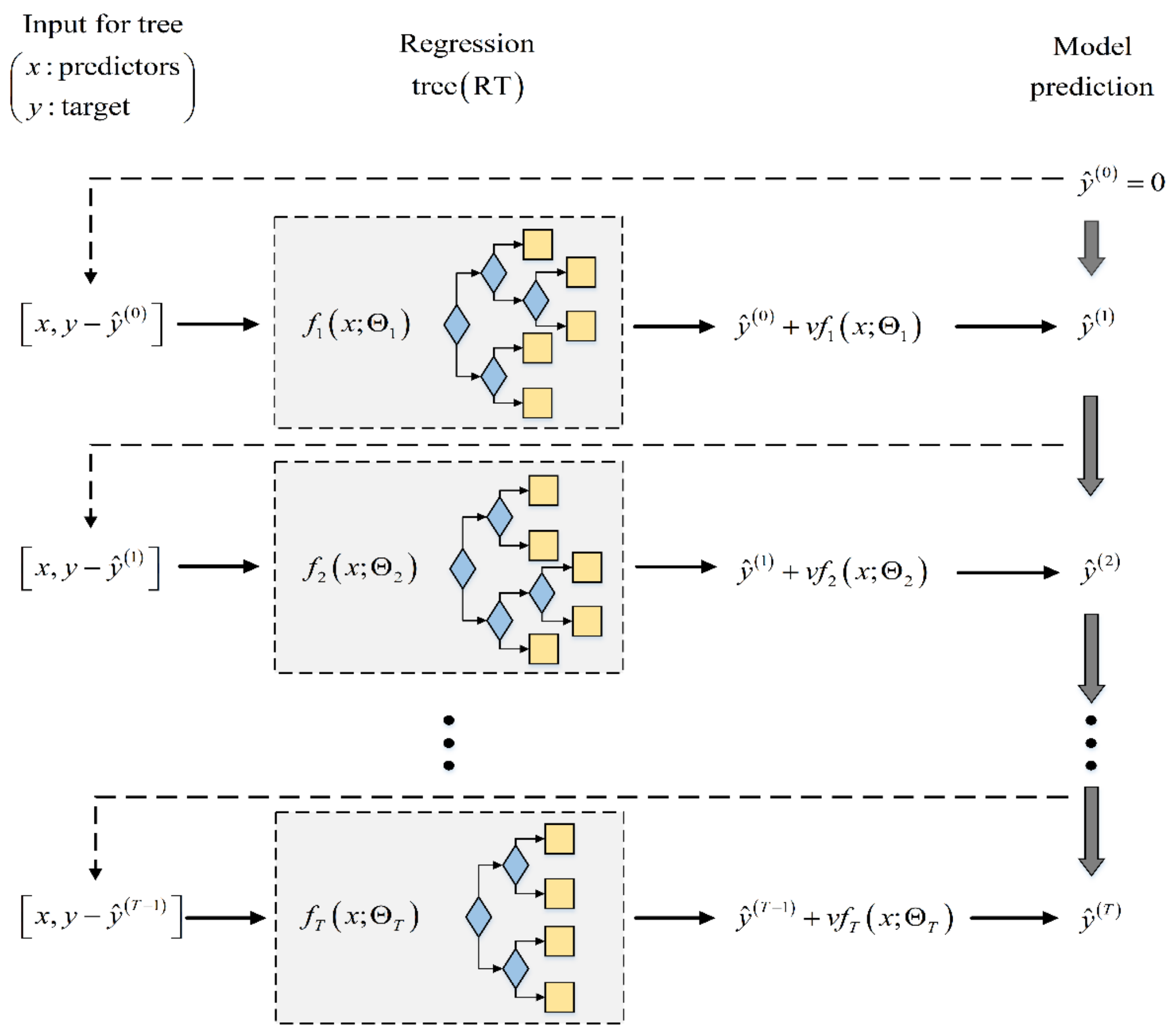

**Figure.2 Principle of XGBoost regression model**

The model of XGboost is as follows, which is shown in Figure 2.

$$\hat{y} = \sum_{k=1}^{K} f_k(x_i), f_k \in F \tag{3.13}$$

The model of loss function is as follows:

$$\mathrm{L} = \sum_{i=1}^{n} l(y_{i,}\hat{y}_i) \tag{3.14}$$

By minimizing the loss function, the function space set is solved, and the regular term is added, the expression is as follows:

$$\mathrm{Obj} = \sum_{i=1}^{n} l(y_{i,}\hat{y}_i) + \sum_{k=1}^{K} \Pi(f_k) \tag{3.15}$$

If the base model is a tree model, $\Pi(f_k)$ represents the complexity of the tree. In order to obtain the optimal solution of the objective function, the forward step-by-step algorithm is adopted, and each step is based on the model set of all previous steps to further optimize the objective function, where the expression of the model at the t th step is as follows:

$$\hat{y}_i^t = \sum_{k=1}^{K} f_k(x_i) = \hat{y}_i^{t-1} + f_t(x_i) \tag{3.16}$$

Where, $g_i$ is the first derivative of the $l(y_i, \hat{y}_i^{t-1})$ function, and $h_i$ is its second derivative. The loss of step t is approximated by the second derivative of the objective function based on the loss of the previous t-1 steps. The objective function is only related to the following formula:

$$\mathrm{Obj} \approx \sum_{i=1}^{n} l\left(g_i f_t(x_i) + \frac{1}{2} h_i f_t(x_i)^2\right) + \Pi(f_k) \quad (3.17)$$

The process of using XGBoost to predict the demand for emergency supplies typically involves the following steps (as shown in Figure 3):

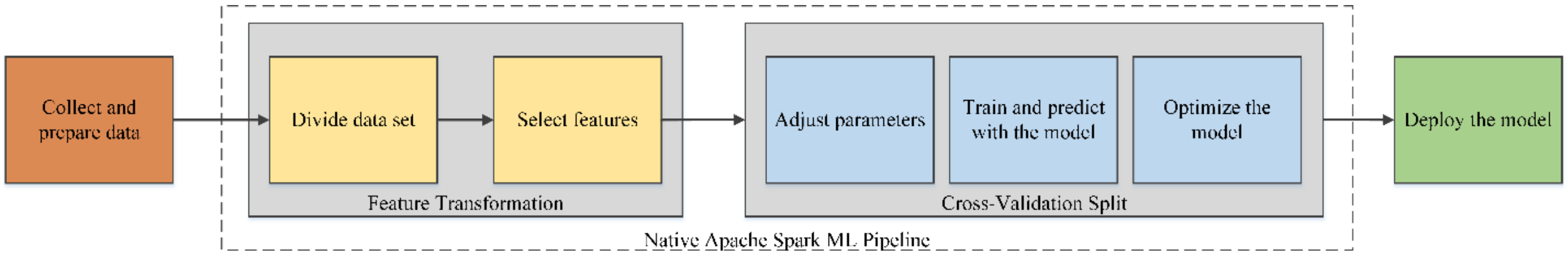

**Figure.3 Steps of XGBoost regression model**

(1) Collect and prepare data: Gather historical data on factors such as demand, time, and location, and perform preprocessing steps such as data cleaning, feature extraction, standardization, and handling of missing values.

(2) Divide data set: Split the data set into training and test sets, typically using 80% of the data for training and 20% for testing.

(3) Select features: Choose features that are relevant to demand, such as time, location, and weather.

(4) Adjust parameters: XGBoost has several adjustable parameters, such as tree depth, learning rate, and regularization terms, which require tuning through techniques like cross-validation to achieve optimal performance.

(5) Train and predict with the model: Use the training set to train the model and then use the test set for prediction, evaluating the accuracy and performance of the model.

(6) Optimize the model: Refine the model based on the prediction results, for example by adjusting parameters or adding new features.

(7) Deploy the model: Apply the trained model to real-world scenarios to predict demand for emergency supplies and provide timely and accurate supply allocation.

### 3.1.3 Material classification

Intuitionistic Fuzzy Theory is a mathematical technique that addresses uncertainty and ambiguity[40]. It has practical applications in emergency response scenarios, where rescuers can use it to evaluate the similarity and substitutability of different materials during multi-disaster rescue operations, through case analysis and summarization[41]. Here are the specific steps involved(As shown in Figure 4):

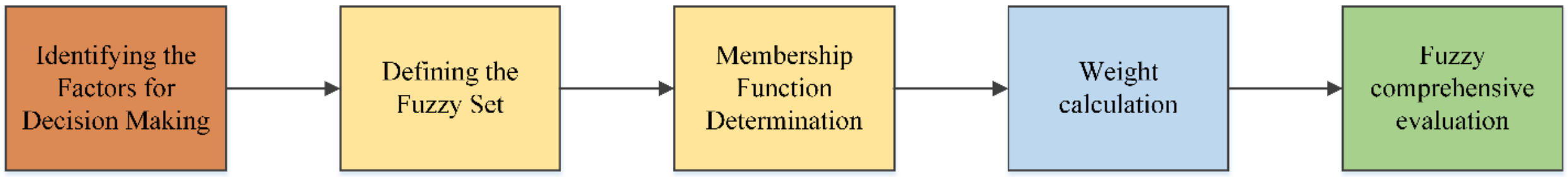

**Figure.4 Steps of Intuitionistic Fuzzy Theory to classify material.**

(1) Identifying the Factors for Decision Making

To begin with, it is essential to identify the factors that are critical for making decisions regarding the selection of an emergency rescue site. These factors may include the number of reserves of rescue supplies, the affected population, the ease of transportation, the surrounding environment, and other relevant factors.

(2) Defining the Fuzzy Set

To prevent the negative impact of feature attribute values with varying dimensions on the calculation of case similarity, it is crucial to normalize the numerical data first. This will aid in defining the fuzzy set.

$$X_{ij} = \frac{x_{ij} - x_j^{min}}{x_j^{max} - x_j^{min}} \tag{3.18}$$

To determine the normalization of decision-making factors, it's crucial to identify the fuzzy set to which each factor belongs. A fuzzy set categorizes possible values in a variable based on specific rules to create a set. As an instance, transportation convenience levels can be divided into four categories: "very convenient," "relatively convenient," "average," and "inconvenient," as follows:

Very convenient - (0, 0, 0.2)

Convenience - (0.12, 0.32, 0.52)

Fair - (0.44, 0.64, 0.84)

Inconvenience - (0.76, 0.96, 1.16, 1.16)

(3) Membership Function Determination

A membership function must be established for each fuzzy set, which represents the degree of membership when the variable takes different values. For example, to determine the degree of membership and non-membership for the decision-making factor of traffic convenience, a triangular membership function can be employed:

$$\vartheta_A(\mathrm{x}) = \begin{cases} \left(1-\frac{\mathrm{x}-a}{\mathrm{b}-a}\right) * D; a < x < b \\ \left(1-\frac{\mathrm{c}-x}{c-b}\right) * D\ ; b \le x < c \\ 1\ ;\quad in\ other \end{cases} \tag{3.19}$$

$$\vartheta_A(\mathrm{x}) = \begin{cases} \left(1-\frac{\mathrm{x}-a}{\mathrm{b}-a}\right) * D; a < x < b \\ \left(1-\frac{\mathrm{c}-x}{c-b}\right) * D\ ; b \le x < c \\ 1\ ;\quad in\ other \end{cases} \tag{3.20}$$

Fuzzy logic operations and fuzzy mathematics methods are used to determine the similarity of each site selection scheme, where a, b, and c represent the vertices of triangular fuzzy numbers and d is a coefficient. This will aid in the calculation of weights for each scheme.

$$\mathrm{Sim}(\mathrm{A},\mathrm{B}) = \left(\frac{\sum_{i=1}^{n} w_i(\min(\mu_A(f_i),\mu_B(f_i)))}{\sum_{i=1}^{n} w_i(\max(\mu_A(f_i),\mu_B(f_i)))} + \frac{\sum_{i=1}^{n} w_i(\min(\vartheta_A(f_i),\vartheta_B(f_i)))}{\sum_{i=1}^{n} w_i(\max(\vartheta_A(f_i),\vartheta_B(f_i)))}\right)/2 \tag{3.21}$$

(4) Weight calculation

Weight is a quantitative value used to describe the degree of influence of each characteristic factor on rescue site selection. Emergency rescuers can assign weights to each characteristic factor through expert interviews, questionnaires, etc. . The weight is usually represented by a value between 0 and 1, and the sum of the weight values of each characteristic factor is equal to 1. Different decision-making factors have different influences on rescue site selection, and emergency rescuers need to determine the weight of each decision-making factor.

The weight $w_i{}^*$ determined by the degree of dispersion of attribute values is obtained by the following formula:

$$w_i{}^* = \frac{\mathrm{s}(\mathrm{x}_i)}{\sum_{m=1}^{M} s(\mathrm{x}_i)} \tag{3.22}$$

(5) Fuzzy comprehensive evaluation

After determining the fuzzy set and membership function of each decision-making factor, it is necessary to carry out fuzzy comprehensive evaluation on these factors to obtain the comprehensive evaluation value of each site selection scheme.

$$\text{Comprehensive evaluation value} = \sum(w_i{}^* \times \mathrm{Sim}(\mathrm{A},\mathrm{B})) \tag{3.23}$$

### 3.2 Multi-objective model

### 3.2.1 Problem description

This study presents a multi-objective mathematical programming model that incorporates the quantity, location, and allocation of joint suppliers under complex disaster scenarios. The objectives of the model include minimizing the cost of emergency resource allocation, maximizing the coverage of joint suppliers, minimizing the gap between demand and suppliers, and minimizing rescue time. The Multi-objective model is shown in Figure 5.

We consider the joint suppliers for multiple disasters at the same time with different types of materials and facilities. The joint suppliers ensure that government owned suppliers, enterprises and existing suppliers coordinate the material reserves. The rescue agencies provide specific and general emergency relief materials. In this approach, companion and converted area of materials are considered.

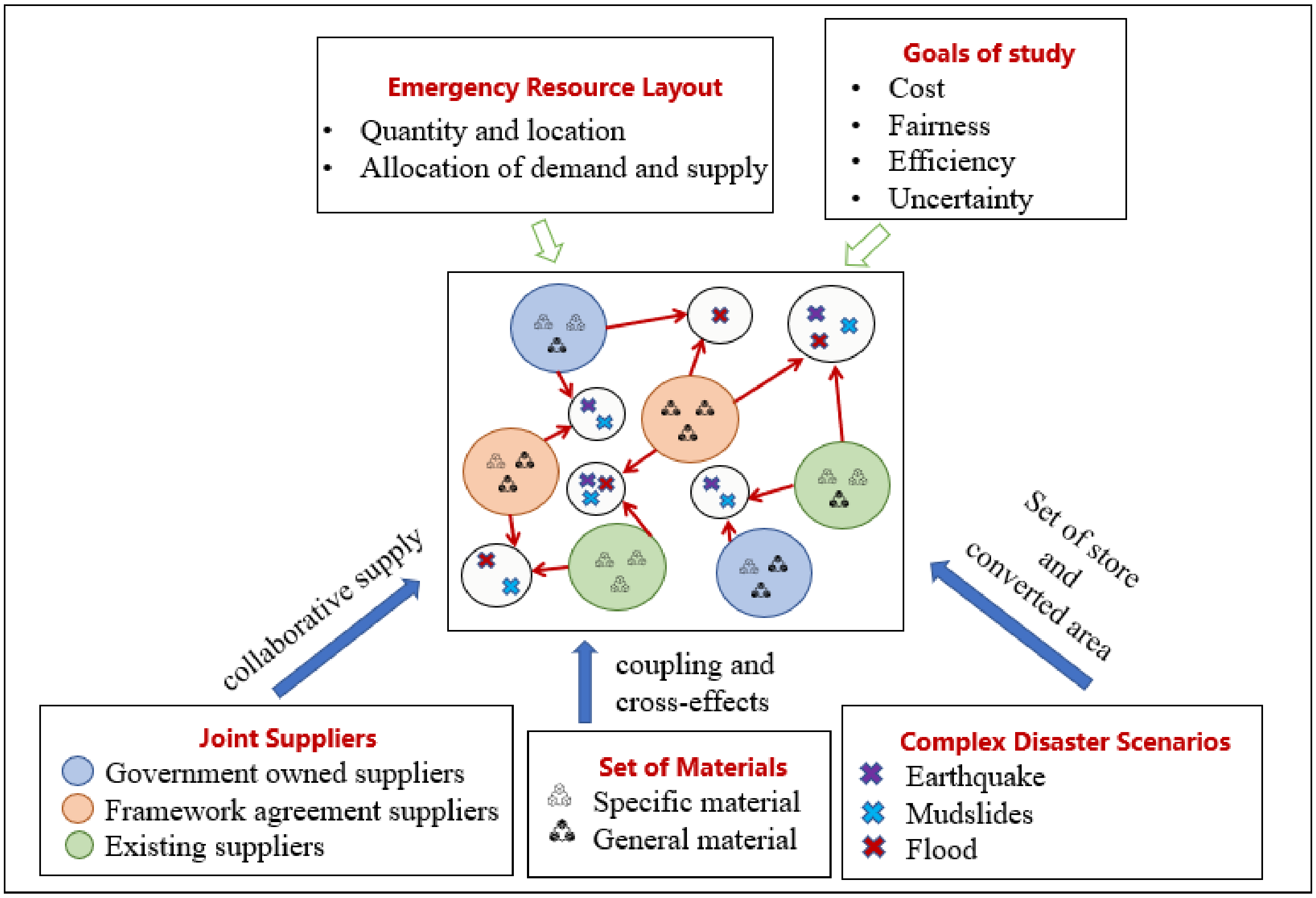

**Figure.5 Framework of Multi-objective model**

### 3.2.2 Model assumptions

The multi-objective model is constructed with the following assumptions:

(1) Assume that the demand of different materials is given by the decision maker.

(2) The demand of each disaster area is random, and at least one supply point is served for it.

(3) The urgency of disaster area obeys a non-convex, non-concave and non-increasing function.

(4) The payment method of the framework agreement suppliers is measured in units of materials.

(5) The constructing and operating costs of existing reserves are not considered.

Considering these assumptions, four objectives are defined for the model including costs, fairness, emergency efficiency and uncertainty.

### 3.2.3 Symbol description

The sets, parameters and variables of the model are introduced as follows:

**Table. 1 Sets of Multi-objective model**

| Name of sets | Description of sets |
|---|---|
| F | A set of various disaster points, fi refers to the type of disasters. $f \in F$, F=f1+f2+f3 |
| M | A set of special materials required for various disasters, mi refers to the type of special materials. $m \in M$, M=m1+m2+m3 |
| C | A set of common supplies required for various disasters, ci refers to the type of common supplies. $c \in C$, C=c1+c2+c3 |
| I | A set of all materials needed for various disasters, $i \in I$, I=M+C |
| W | A set of government owned suppliers, $w \in W$ |
| P | A set of framework agreement suppliers, $p \in P$ |
| K | A set of existing suppliers, $k \in K$ |
| J | A set of alternative suppliers, $j \in J$, J=W+P+K |

**Table. 2 Parameters of Multi-objective model**

| Name of Parameters | Description of sets |
|---|---|
| $N_f$ | The number of disaster area f, $f \in F$ |
| $N_i$ | The number of types of material i, $i \in I$ |
| $N_w$ | The number of types of government owned supplier w, $w \in W$ |
| $N_p$ | The number of types of framework agreement supplier p, p∈P |
| $N_k$ | The number of types of existing supplier k, k∈K |
| $N_j$ | The number of supplier j, j∈J, J=W+P+K |
| $V_w$ | The maximum area of the government owned supplier w, w∈W |
| $V_p$ | The maximum area of the framework agreement supplier p, p∈P |
| $V_k$ | The maximum area of the existing supplier k, k∈K |
| $D_{1i}$ | General material demand in the disaster area, i∈C |
| $\alpha_i$ | Conversion coefficient of general material demand, i∈C |
| $D_{2i}$ | Specific material demand in the disaster area, i∈M |
| $S_{ki}$ | The amount of material i provided by existing suppliers k, i∈I, k∈K |
| $B_w$ | The constructing and operating costs of the government owned supplier w, w∈W |
| $B_{pi}$ | The unit cost of material i provided by framework agreement supplier p, i∈I |
| $B_{ki}$ | The unit cost of material i provided by existing supplier k, i∈I |
| $G_{wi}$ | Cost of unit quantity of material I provided by government owned suppliers w, i∈I, w∈W |
| $Q_i$ | The converted area of material i, i∈I |
| $N_w^{max}$ | The maximum number of government owned supplier w, w ∈ W |
| $N_w^{min}$ | The minimum number of government owned supplier w, w ∈ W |
| $N_p^{max}$ | The maximum number of framework agreement supplier p, p∈P, i∈I |
| $N_p^{min}$ | The minimum number of framework agreement supplier p, p∈P |
| $N_k^{max}$ | The maximum number of existing supplier k, k∈K, i∈I |
| $N_k^{min}$ | The minimum number of existing supplier k, k∈K |
| $E_j$ | The ratio of the area covered by radius of the alternative suppliers j to the total area of the area , j∈W∪P∪K |
| $C_{(djf)}$ | The degree of satisfaction covered by supplier j in disaster area f, j∈W∪P∪K,f∈F |
| $d_{jf}$ | The distance from supplier j to disaster area f, j∈W∪P∪K,f∈F |
| $V_{jf}$ | The velocity from supplier j to disaster area f, j∈W∪P∪K,f∈F |
| $R_{jf}$ | The maximum radius from supplier j to disaster area f is determined by the maximum rescue time, $j \in W \cup P \cup K, f \in F$ |
| $r_{jf}$ | The minimum radius from suppliers j to disaster area f, which ensures that disaster area is covered by at least one supplier, $j \in W \cup P \cup K$, $f \in F$ |

**Table. 3 Decision Variables of Multi-objective model**

| Name of Variables | Description of Decision Variables |
|---|---|
| $X_j$ | If the alternative supplier j is selected, take 1, otherwise take 0, j∈W⋃P |
| $y_{jf}$ | If the selected supplier j serves the disaster area f, take 1, otherwise take 0, f∈F, j∈W⋃P⋃K |
| $Z_{wi}$ | The amount of material i provided by government owned supplier w, w∈W, i∈I |
| $Z_{pi}$ | The amount of material i provided by framework agreement supplier p, p∈P, i∈I |
| $Z_{ki}$ | The amount of material i provided by existing supplier k, k∈K, i∈I |

### 3.2.4 Mathematical modelling

The multi-objective model of joint suppliers for emergency resource layout under complex disaster scenarios is as follows:

**objectives.**

(1) Minimizes total cost of emergency resource layout

$$Min1 = \sum_{w \in W} \left(Bw + \sum_{i \in I} (Gwi \times Zwi \times Xw)\right) + \sum_k Z_{ki} + \sum_{p \in P} \left(\sum_{i \in I} (Xp \times Bpi)\right) \quad (3.1)$$

In the selection of emergency resource suppliers, cost control is an important reference target due to the limitation of actual economic capacity. Therefore, in the design of location scheme for emergency resource supplier, the use efficiency of funds should be improved and the cost should be reduced as far as possible in the case of meeting the needs. The model of the total cost of the cost target is to minimize the emergency resources, which is the first objective function of the model. The cost of emergency resource layout mainly includes three parts. The first part is related to constructing and operating costs of the government owned suppliers. The second part of equation (1) implies the cost of materials provided by government owned suppliers and framework agreement suppliers.

(2) Maximizes coverage expectation of alternative suppliers

$$Max3 = \sum_{j \in W \cup P \cup K} \left(E_j \times y_{jf} \times F(d_{jf})\right) \quad (3.2)$$

To ensure that all demand areas can receive corresponding emergency resources when facing disaster threats, ensuring the fairness of emergency rescue is also an important goal of the problem of emergency resource location selection. Here, the emergency resource utility of each demand area is defined as the degree of satisfaction and coverage of disaster area by suppliers. It consists of two parts; the first part is the ratio of the area covered by radius of the alternative suppliers j to the total area of the area. The second part refers to the degree of satisfaction covered by supplier j in disaster area f, which is formulated as follows:

$$F(d_{jf}) = \begin{cases} 1, d_{jf} \leq r_{jf} \\ 1 - \frac{d_{jf} - r_{jf}}{R_{jf} - r_{jf}}, r_{jf} \leq d_{jf} \leq R_{jf} \\ 0, d_{jf} \geq R_{jf} \end{cases} \tag{3.3}$$

(3) Minimizes the difference between demand and suppliers.

$$\text{Min3} = (\sum_{i \in C} (\sum_k (S_{ki} + Z_{ki}) + \sum_w Z_{wi} + \sum_p Z_{pi} - \sum_f (D1_{fi} * \alpha_i)) + (\sum_{i \in M} (\sum_k (S_{ki} + Z_{ki}) + \sum_w Z_{wi} + \sum_p Z_{pi} - \sum_f D2_{fm})) \tag{3.4}$$

The third objective function, which is shown in equation above, minimizes the difference between demand of disaster areas and materials provided by existing suppliers, government owned suppliers and framework agreement suppliers. In addition, demand of disaster areas is consisting of general and specific materials.

(4) Minimizes rescue time

$$Min4 = \sum_{j \in W \cup P \cup K} \left( y_{jf} \times \frac{d_{jf}}{V_{jf}} \right) \tag{3.5}$$

Considering that the important purpose of emergency resource reserve is to save people's lives, and the effectiveness of emergency resources is directly related to the main goal of saving lives. The effectiveness of emergency reserve resources is mainly determined by its timeliness. Therefore, the fourth objective function minimize sum of the rescue time from suppliers to disaster areas.

s.t.

$$\mathrm{N}_{\mathrm{pi}}^{\mathrm{min}} \leq Z_{pi} \leq \mathrm{N}_{\mathrm{pi}}^{\mathrm{max}} \tag{3.6}$$

$$\mathrm{N}_{\mathrm{w}}^{\mathrm{max}} \geq \sum_w X_w \geq \mathrm{N}_{\mathrm{w}}^{\mathrm{min}} \tag{3.7}$$

$$N_p^{max} \geq \sum_p X_p \quad \geq N_p^{min} \tag{3.8}$$

$$N_k^{max} \geq \sum_k X_k \quad \geq N_k^{min} \tag{3.9}$$

$$\sum_i Q_i \quad \times Z_{wi} \leq Vw \tag{3.10}$$

$$\sum_i Q_i \times Z_{ki} \leq V_k \tag{3.11}$$

$$\sum_i Q_i \times Z_{pi} \leq V_p \tag{3.12}$$

$$\sum_{j \in W_\cup P_\cup K} Z_{jm} \quad \leq D1_{fc} \times \alpha_c \tag{3.13}$$

$$\sum_{j \in W_\cup P_\cup K} Z_{jm} \quad \leq D2_{fm} \tag{3.14}$$

$$Z_{jm} \leq M \sum_{f1} y_{jf} \quad (m \in m1 \cup m2 \cup m3 \tag{3.15}$$

$$y_{jf} \leq Xj, \mathrm{j} \in \mathrm{W \cup P \cup K}, \forall \mathrm{f} \in \mathrm{F} \tag{3.16}$$

$$X_j \leq \sum_i Z_{ij}\ , \ \forall j \in W \cup P \cup K \tag{3.17}$$

$$y_{wf} + y_{kf} + y_{pf} \geq 1, \mathrm{f} \in \mathrm{F} \tag{3.18}$$

$$y_{jf}, Xj \in \{0,1\} \tag{3.19}$$

$$\mathrm{Zwi, Zpi, Zki} \geq 0 \tag{3.20}$$

Constraint (3.6) is the upper and lower bounds the quantitative restrictions of framework agreement suppliers. Constraint (3.7) -(3.9) represents that the upper and lower bounds of suppliers. Constraint (3.10) -(3.12) controls that the storage capacity of suppliers by the area. Constraint (3.13) -(3.14) is specified for the quantity of each material greater than the demand. Constraints (3.15) defines that if the specific material m1 is dedicated to the supplier j, then j serves at least one demand area f. Constraint (3.16) expresses that only when supplier j is selected to store materials, it is able to serve demand area f. Constraint (3.17) implies that only when supplier j is selected, it is able to store a certain number of reserves. Constraint (3.18) ensures that at least one supplier j serves disaster area f. Constraint (3.19) indicates the binary variables. And ultimately, constraint (3.20) are non-negativity constraints.

### 3.3 Solution approach

Due to the multi-objective nature of the present model, optimizing one objective function may have a negative impact on other objective functions[42]. Therefore, multi-objective optimization methods are necessary to obtain a set of solutions that optimize all objective functions as much as possible without degrading any one objective function[43]. In a multi-objective optimization problem, the Pareto dominance relation defines a good solution[44]. One solution dominates another if it outperforms it in all objective functions, and two solutions are non-dominated if neither dominates the other. A Pareto optimal solution is one in which no solution can dominate it.

NSGA-II is a genetic algorithm-based multi-objective optimization algorithm that can effectively handle high-dimensional, non-convex, and nonlinear problems and can generate uniformly distributed non-dominated solution sets[45, 46]. The NSGA-II algorithm involves several steps shown in Figure 6.

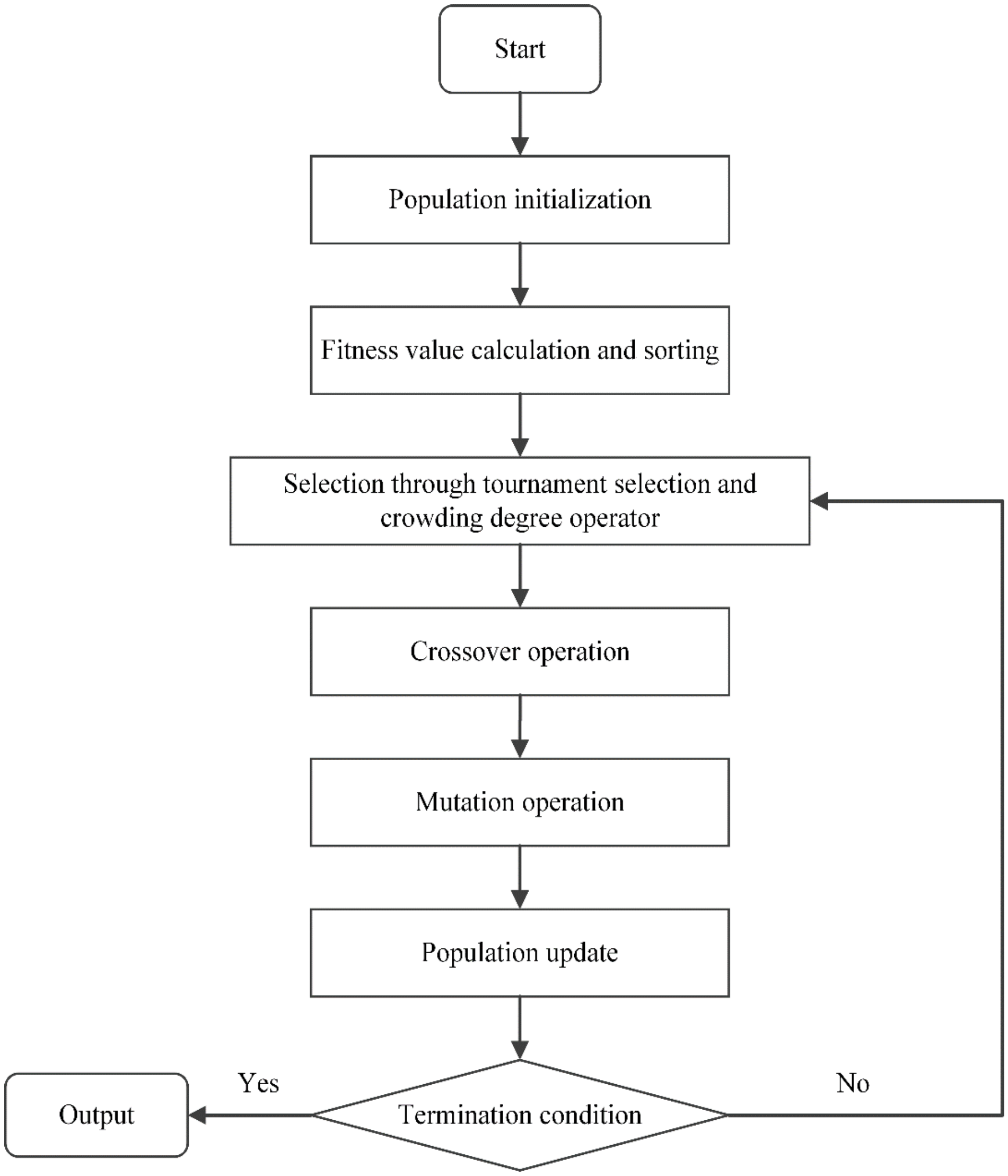

**Figure. 6 Framework of NSGA-II model**

The multi-objective emergency rescue site selection model is a model that selects the optimal rescue site selection scheme during the emergency rescue process based on various factors and indicators. NSGA-II is a commonly used multi-objective optimization algorithm for optimizing multiple objective functions[47]. In the NSGA-II algorithm for the multi-objective emergency rescue site selection model, the key operations include encoding the candidate rescue site selection schemes into a set of decision variables, randomly generating an initial population, selecting individuals using the non-dominated sorting and crowding distance strategy, performing crossover and mutation operations, and performing non-dominated sorting and crowding distance calculation to ensure population diversity and distribution[48]. The algorithm is terminated when the preset maximum number of iterations is reached or the population's diversity and distribution meet the requirements, and the optimal solution is output.

# 4. Numerical study

To verify the proposed multi-objective model and its solution algorithm, the following calculation examples are selected for verification. This chapter begins by analyzing an example to validate the algorithm's rationality and applicability based on its research background. The first step involves using the XGboost integrated learning method to predict the demand for disaster materials and verifying the case. In the next step, the demand is determined in complex disaster scenarios by utilizing the multi-disaster coupling degree and the multi-material alternative conversion coefficient. Finally, a multi-objective decision-making model is established to validate the pre-storage scheme using the proposed NSAG-II-XGB solution algorithm. Several calculation examples are chosen for verification.

## 4.1 Data description and parameter settings

To verify the proposed multi-objective model and its solution algorithm, the following calculation examples are selected for verification. The validity of our model is demonstrated in this study using relevant data from a province that is susceptible to disasters. To account for limitations in obtaining real data and to focus on a specific scope, we examined three types of disasters: earthquakes, floods and mudslides. In addition, we took into consideration the interdependent and combined effects of multiple disasters. This chapter begins by analyzing an example to validate the algorithm's rationality and applicability based on its research background. The first step involves using the XGboost integrated learning method to predict the demand for disaster materials and verifying the case. In the next step, the demand is determined in complex disaster scenarios by utilizing the multi-disaster coupling degree and the multi-material alternative conversion coefficient. Finally, a multi-objective decision-making model is established to validate the pre-storage scheme using the proposed NSAG-II-XGB solution algorithm. Several calculation examples are chosen for verification.

### 4.1.1 Disaster data

The data source for this article is the disaster data of Sichuan Province, located in Southwest China, from August 23, 2002 to December 30, 2022. The province is prone to natural disasters due to its geographical location and terrain conditions, and as a result, the Sichuan Provincial Government and relevant agencies have established a comprehensive disaster point data system to effectively monitor and respond to various natural disasters. The article documents the essential information of disasters that have occurred in various locations in Sichuan Province, including specific location, magnitude, focal depth, and occurrence time. Additionally, the article includes fundamental information on the type and severity of disasters, as well as the specific location, scale, and shape of debris flow disasters. Figure 7 illustrates these details.

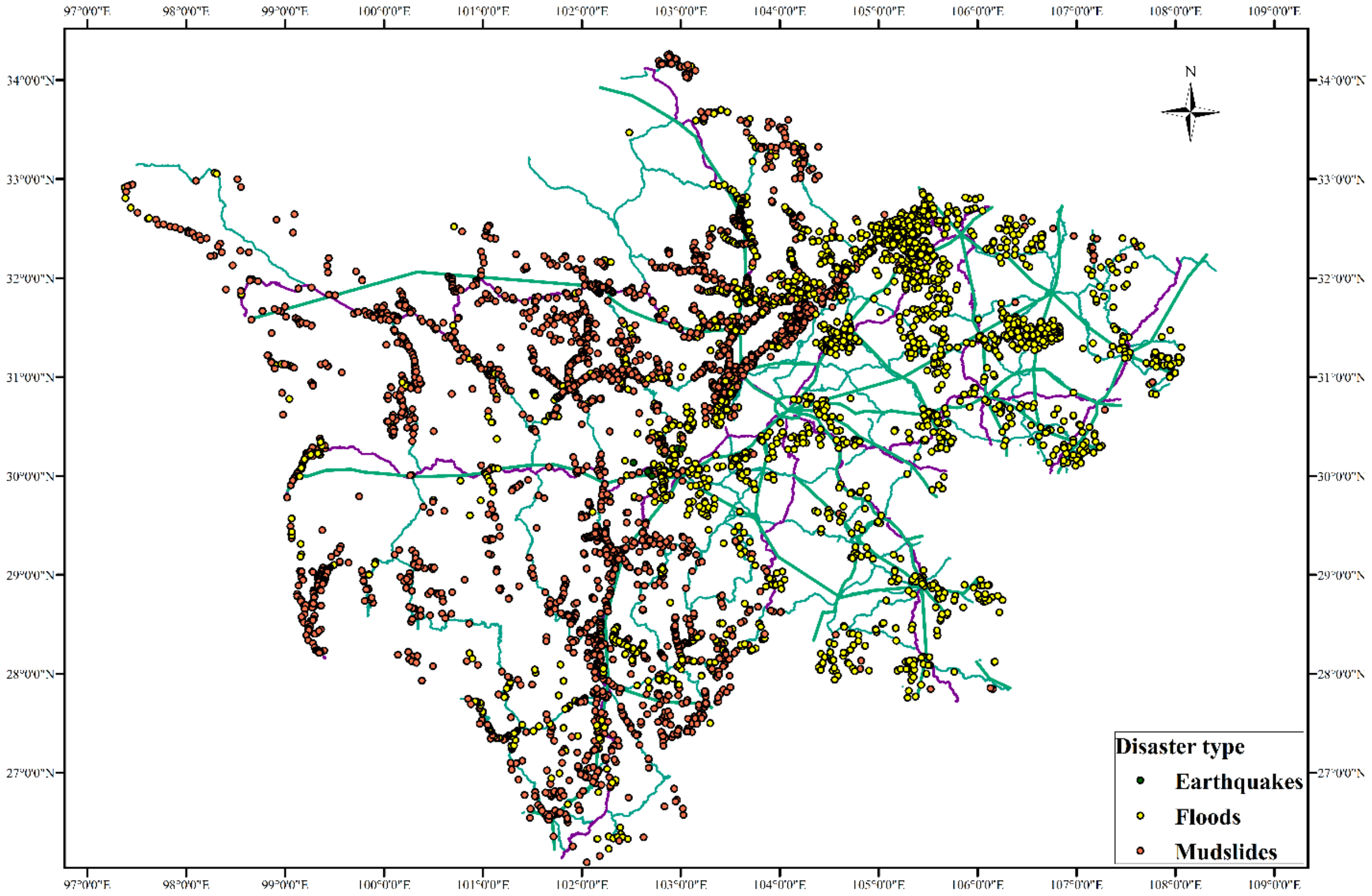

**Figure. 7 Disaster point collection**

Following the data preprocessing explained in earlier chapters, the training and testing datasets have been created. Table 4 displays the definitions for the disaster point fields.

**Table. 4 Disaster point field meaning table**

| Number | Description | Type |
|---|---|---|
| 1 | Disaster body type | string |
| 2 | Geographic Location | string |
| 3 | Longitude | float |
| 4 | Latitude | float |
| 5 | Seismic intensity | float |
| 6 | Threat population | int |
| 7 | Threaten property | int |
| 8 | Disaster level | int |
| 9 | Danger level | int |
| 10 | Prevention Suggestions | string |
| 11 | Monitoring recommendations | string |
| 12 | Destroy houses | int |

This study utilizes data from disaster-prone provinces to validate the efficacy of the model. Due to limitations in acquiring actual data and the research scope, the paper focuses on three types of disasters: earthquake, flood, and debris flow. Furthermore, it includes 22 relevant materials in the disaster area, taking into account the potential coupling and cross-effects of multiple disasters, as depicted in Table 5.

**Table. 5 Field meaning of 4 Rescue point**

| | Intensity | Population | Property | Level of Disaster |
|---|---|---|---|---|
| Intensity | 1 | 3 | 1 | 0.33 |
| Population | 0.33 | 1 | 0.50 | 0.20 |
| Property | 1 | 2 | 1 | 0.33 |
| Level of Disaster | 3 | 5 | 3 | 1 |

Table 6 displays the risk index and coupling degree of the seven disaster scenarios, which were calculated using the risk assessment method for complex disaster scenarios proposed earlier, based on historical demand data from disaster sites.

**Table. 6 Calculation table of disaster coupling and risk index**

| Coupling | Type of disaster | Coupling index | Risk index |
|---|---|---|---|
| Coupling of double disasters | Floods and mudslides | 0.499 | 8.58 |
| | Floods and Earthquakes | 0.498 | 8.31 |
| Coupling of three disasters | Earthquakes and mudslides | 0.286 | 7.97 |
| | Earthquakes, floods and mudslides | 0.283 | 7.73 |

### 4.1.2 Supply data

There are three types of supply points: the provincial government supply point, the contracted enterprise reserve point, and the existing supplier. The rescue response point is an established infrastructure that provides rescue services and support during emergencies. Governments, non-governmental organizations, or private institutions usually establish these response points to ensure that support and assistance can be quickly and effectively provided in emergency situations. Figure 8 illustrates the distribution of these points.

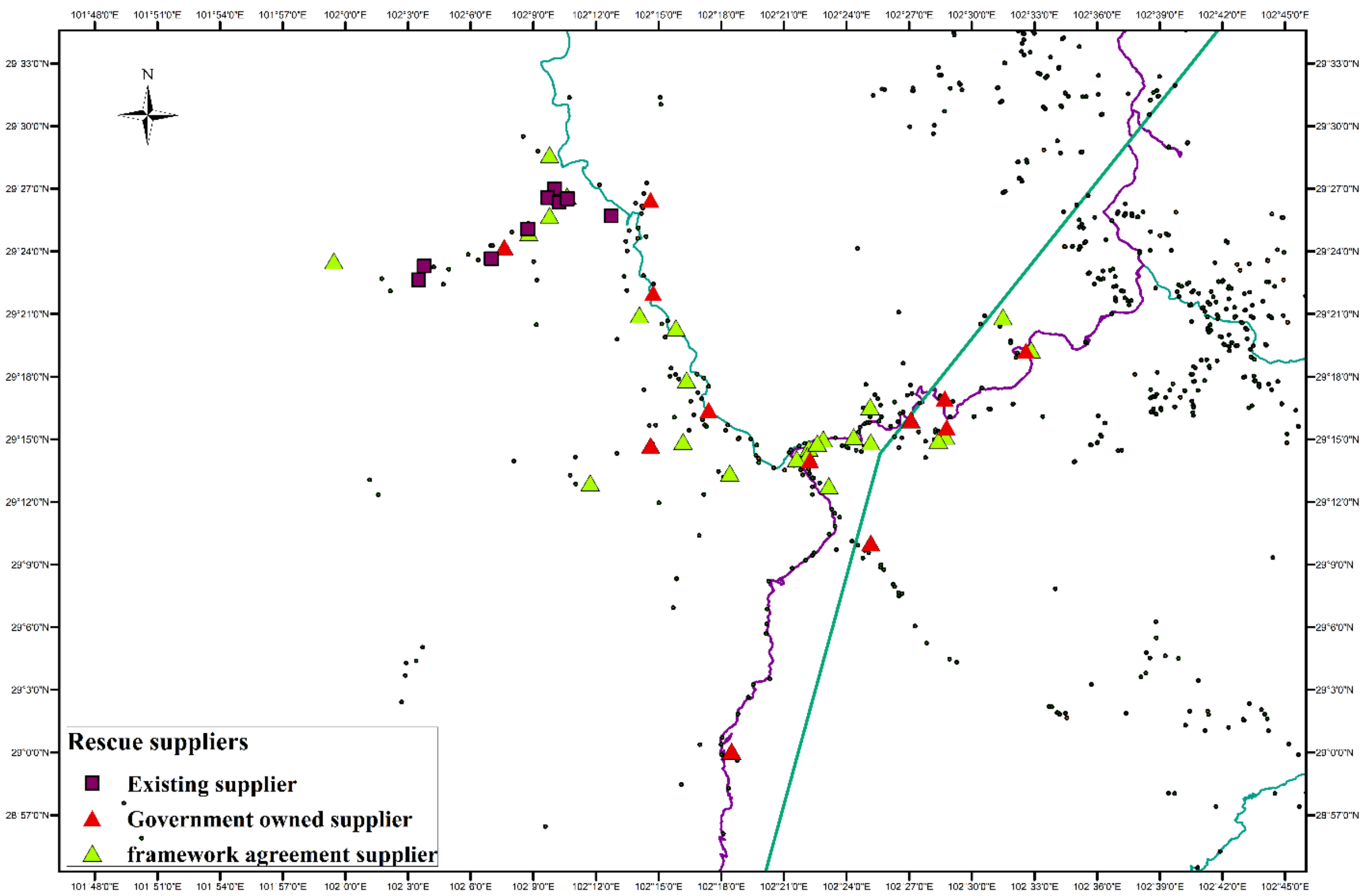

**Figure.8 Distribution of alternative suppliers**

The supply point data includes the material reserve of the material supply point, the type of supply point, the location of the supply point, the volume limit, etc. The meaning of the fields of the supply point is shown in Table 7.

**Table. 7 Meaning table of rescue point fields**

| Number | Description | Type |
|---|---|---|
| 1 | Material type | string |
| 2 | Specific materials | string |
| 3 | Material quantity | int |
| 4 | Location | string |
| 5 | Supplier type | string |
| 6 | Supplier volume | float |
| 7 | Longest rescue time | int |
| 8 | Longitude | float |
| 9 | Latitude | float |

Various major materials are categorized based on their urgency for emergency rescue and the specific types, along with the longest response time, are identified according to [49].

**Table. 8 The longest timetable for the rescue of main materials after the disaster**

| Material type | Specific type | Maximum rescue time (h) |
|---|---|---|
| Special Materials | Rescue equipment | 0.5 |
| | Medical supplies | 1 |
| General supplies | Emergency Food and Daily Supplies | 3 |
| | Tents and other logistical supplies | 2 |

In this paper, 22 types of materials are classified as either general or special. They are then converted based on their emergency classification and substitutability. The area occupied by each material is determined by its supporting conditions. Using historical disaster data, the demand for materials is predicted using XGBOOST. While there may be some approximation errors in demand data, they are not a major concern given the scope and purpose of the study. The government suppliers, agreed suppliers, and existing suppliers provide information on the existing material reserves, supply costs, and demand in the disaster area. The reserve situation determines the constraints on material supply, as shown in Table 8.

This paper focuses on an area that is susceptible to natural disasters like earthquakes, mudslides, and landslides. The study analyzes the supply situation in disaster-prone areas based on factors such as disaster points, supply points, and types of supply points. To determine the satisfaction level of alternative suppliers and the proportion of the radius coverage area, the distance, maximum, and minimum radius from supplier j to disaster area f are calculated using ARCGIS. The calculation method considers the speed limit of different types of roads (as detailed in Table 9 and Figure 8).

**Table. 9 Matching fields of supply points and disaster points**

| Number | Description | Type |
|---|---|---|
| 1 | Disaster point number | int |
| 2 | Rescue point number | int |
| 3 | Distance from rescue point to disaster point | int |
| 4 | Minimum coverage radius from rescue point to disaster point | float |
| 5 | Maximum coverage radius from rescue point to disaster point | float |
| 6 | Coverage expectation from rescue point to disaster point | float |
| 7 | Rescue time from rescue point to disaster point | float |
| 8 | Rescue speed from rescue point to disaster point | float |

The study involves 22 types of general and special materials. The government-owned suppliers and framework agreement suppliers provide data on existing material storage, cost of supplies, and demand in disaster areas. The problem involves 3429 decision variables subject to 1068 constraints, and potential errors in demand data approximation are acknowledged but are not expected to be a significant issue given the scope of the study.

### 4.2 Analysis of numerical results

The Pareto solution set is a useful tool for guiding emergency rescue site selection decisions and identifying the best rescue plan. This set comprises solutions that cannot be improved further in a multi-objective optimization problem. In the context of selecting a location for emergency rescue, the Pareto solution set could include various solutions, each representing a possible optimal solution. For instance, in terms of rescue response time and cost, the Pareto solution set may consist of solutions with varying response times and costs. By analyzing the Pareto solution set, decision makers can better balance different objectives and choose the most suitable solution. This method is highly effective in making trade-offs among multiple objectives and selecting the optimal solution. This study employs NSGA-II-XGB to solve the emergency rescue location problem and finds a total of 40 sets of Pareto solutions, as depicted in Figure. 9.

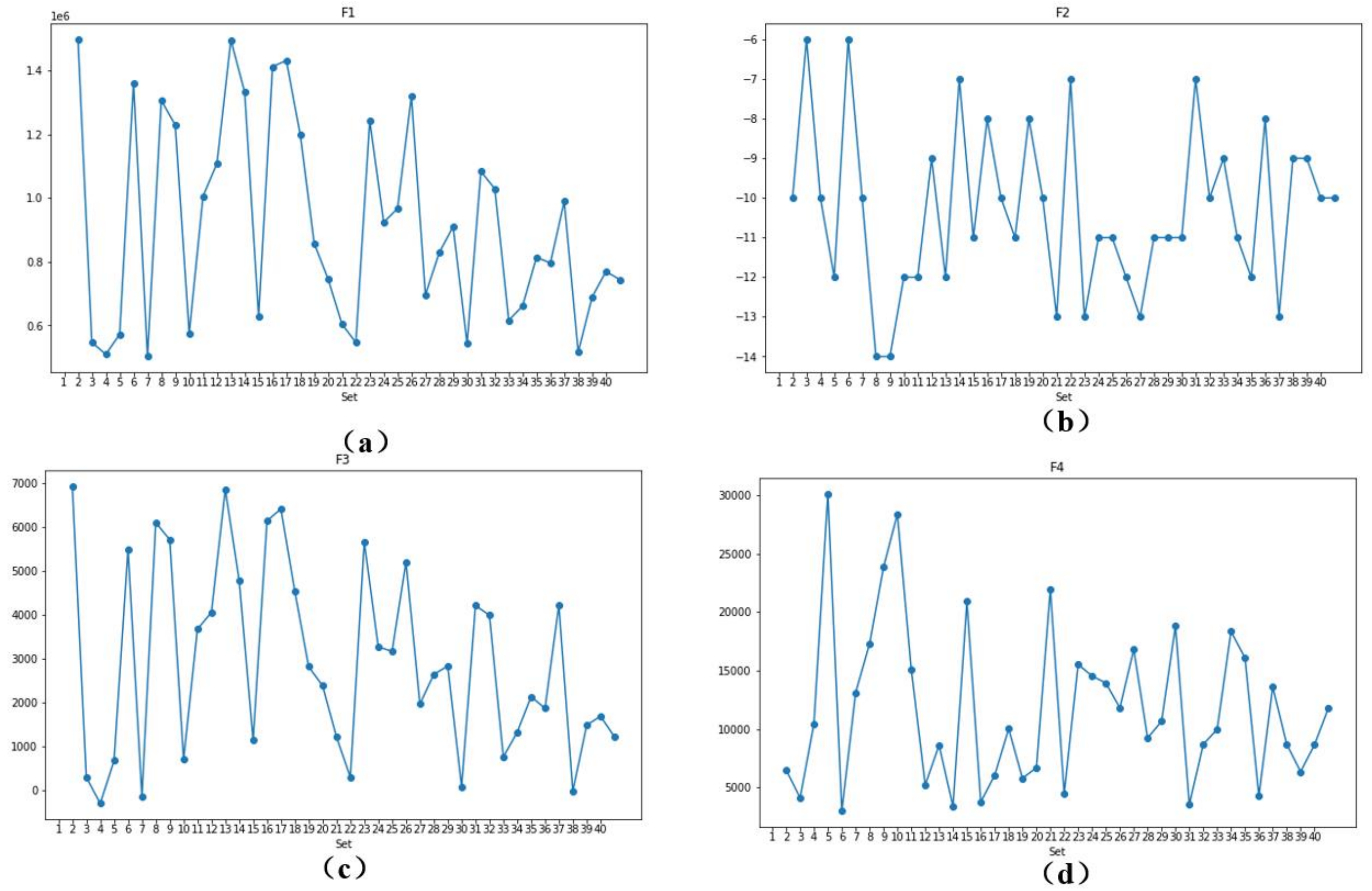

**Figure. 9 Parato solutions of Multi-objective model**

The cost of a rescue operation comprises various elements, such as transportation, maintenance, personnel, and equipment. The optimal selection of a rescue site can minimize these costs, which in turn affect the efficiency and effectiveness of the operation. For instance, choosing a site close to the disaster area with convenient transportation can reduce transportation costs, selecting appropriate equipment and materials can lower maintenance costs, and deploying personnel and resources efficiently can improve overall efficiency and reduce costs. When the priority is to minimize costs, Pareto analysis shows that schemes 2, 3, 4, 6, 9, 21, 29, and 37 provide the most cost-effective solutions, one of the solution with scheme 2 is shown in Figure 10.

When emphasis is given to coverage and fairness, it is crucial to choose a site that covers the maximum number of affected people and distributes rescue resources fairly across different regions. Pareto analysis suggests that schemes 4, 7, 8, 9, 20, 22, 26, 34, and 36 are optimal in terms of coverage and fairness.

Demand forecasting is crucial when attention is paid to the difference between demand and supply. Accurate prediction of rescue site requirements can help relevant organizations to prepare in advance, including the deployment of personnel, materials, and equipment. This can ensure a rapid and timely response in the event of an emergency, reducing casualties and property losses. Pareto analysis indicates that schemes 2, 3, 4, 6, 9, 21, 29, and 37 offer the most accurate prediction.

When time is critical, selecting a rescue site promptly is essential. Delays in site selection can lead to reduced rescue efficiency and increased risk of casualties. Effective measures need to be taken to speed up the response time. Pareto analysis suggests that schemes 2, 5, 11, 13, 15, 30, 35 provide the solutions with the shortest rescue time.

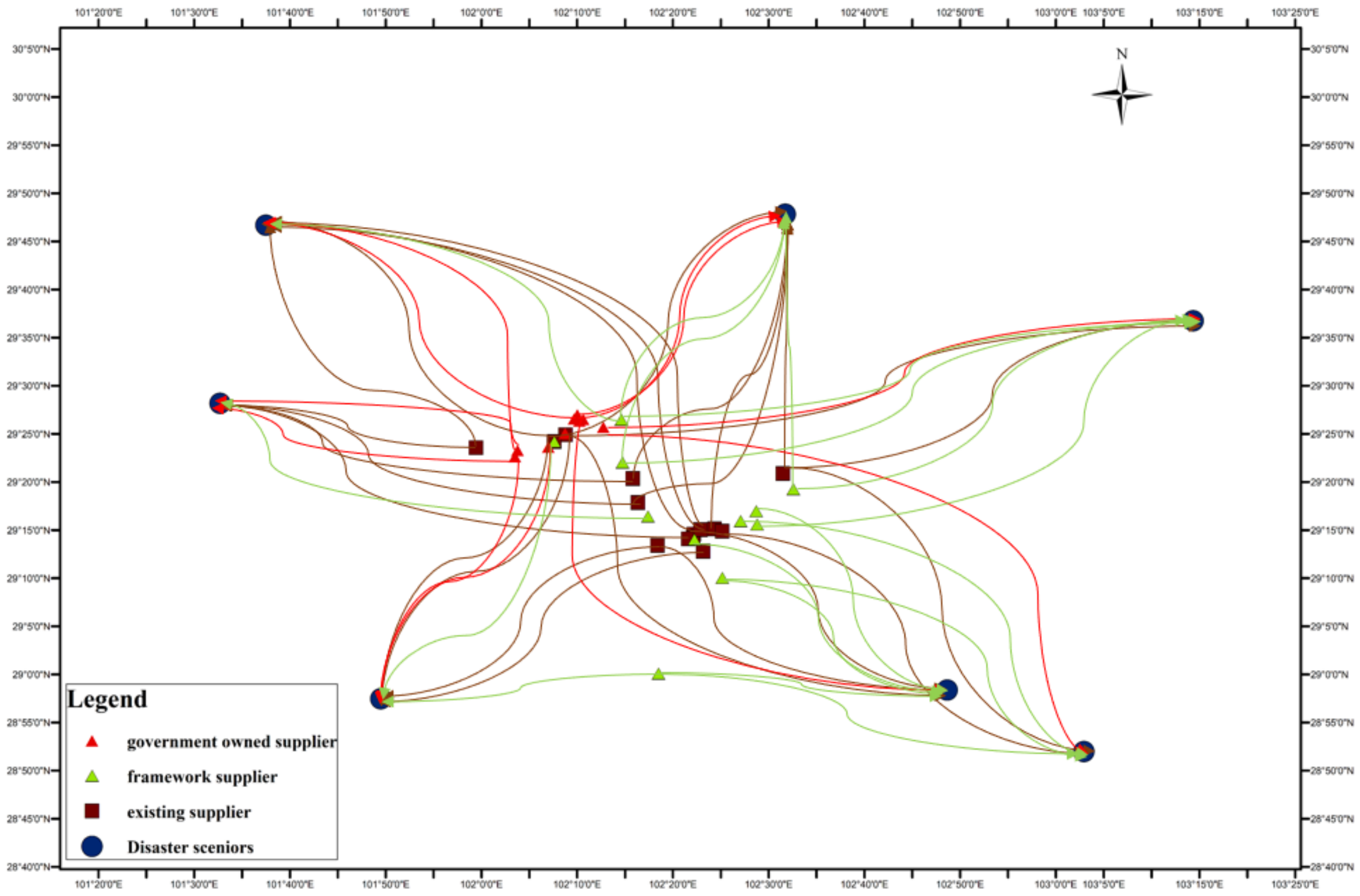

**Figure. 10 Example of Parato solution with lowest cost**

## 5. Conclusion

After complex disasters occur, difficulties in stockpiling materials arise due to high cost, insufficient coverage, and the interaction between different disasters. Relying solely on government or contracted suppliers is not enough to meet the needs of multiple disaster areas and complete the preparation of emergency supplies. Therefore, the selection of agreement suppliers and material reserves is crucial to the layout of emergency resources for multiple disasters. A multi-objective model that comprehensively considers cost, fairness, emergency efficiency, and uncertainty can effectively respond to changes in disaster needs and improve the efficiency of emergency rescue.

To provide reliable strategies for emergency rescue in complex disaster scenarios, this study integrates these factors into a multi-objective model. Using a scenario-based approach to represent complex disasters, a multi-objective mathematical model is built to obtain supplier storage locations and quantities. This model demonstrates a cooperative response between governments, agreements, and existing suppliers for different kinds of disasters. Based on a large dataset of disaster-prone provinces, the primary goal of this paper is to minimize the overall expected cost with government suppliers, contracted suppliers, and existing suppliers. Additionally, this paper considers the supplier's maximum coverage expectation and attempts to use the multi-objective model to reduce the difference in material demand forecasting among various suppliers and the rescue time.

The case analysis shows that the proposed method can estimate the quantity and location of material reserves well and can be applied to various types of disaster areas. Therefore, this paper provides valuable solutions for the location and layout of supplies for government suppliers, contract suppliers, and existing suppliers. The NSGA-II-XGB algorithm is proposed and verified to perform well in dealing with complex disasters.

Compared with previous studies[48, 50, 51], this paper makes several related contributions. Firstly, it considers the coupling interaction of complex disaster scenarios and determines the impact of the interaction of multiple complex disasters on the location selection of emergency rescue. Secondly, it proposes a collaborative response of multiple rescue subjects based on the existing rescue response points. Thirdly, it considers the conversion and substitution of general materials and materials in specific disaster scenarios. Finally, the model proposed in this paper provides a series of available options for decision according to the distribution of Pareto points and the preference of the target.

However, this study is only a preliminary exploration, and future scholars should conduct a more systematic and professional analysis of this topic. One of the future research directions can be to consider different types of vehicles and transportation costs, as well as the routing problem and vehicle scheduling of emergency rescue. Additionally, considering more objective functions and proposing a new meta-heuristic algorithm can also be used as a field of future research.

## Acknowledgment

This research was partially funded by Natural Science Basic Research Plan in Shaanxi Province of China, Transportation-Energy-Environment Complex Mechanism and System Regulation(Grant Number 2021JC-27) , Shaanxi Provincial Key Science and Technology Innovation Group, Digitalization of highway transportation infrastructure and intelligent disaster management and control innovation team(Grant Number 2023-CX-TD-11)